\documentclass[aps,prb,amsfonts,amssymb,twocolumn,amsmath,floatfix,showpacs]{revtex4-1}
\usepackage{amsbsy}
\usepackage{subfigure}
\usepackage[dvips]{color}
\usepackage[dvips]{graphics}
\usepackage{graphicx} % Include figure files
\usepackage{bm} % bold math
\usepackage[T2A]{fontenc}
\usepackage[cp1251]{inputenc}
\usepackage[english]{babel}

\newcommand{\be}{\begin{equation}}
\newcommand{\ee}{\end{equation}}
\newcommand{\bel}{\begin{align}}
\newcommand{\eel}{\end{align}}
\newcommand{\bem}{\begin{multline}}
\newcommand{\eem}{\end{multline}}
\newcommand{\beq}{\begin{equation}}
\newcommand{\eeq}{\end{equation}}
\newcommand{\bea}{\begin{eqnarray}}
\newcommand{\eea}{\end{eqnarray}}

\newcommand{\arccot}{\mathrm{arccot}\,}

\DeclareMathOperator{\sgn}{sgn}

\begin{document}

\title{Proximity-induced minimum radius of superconducting thin rings\\ closed by the Josephson
$0\,$ or $\pi\,$ junction}

\author{Yu.\,S.~Barash}

\affiliation{Institute of Solid State Physics, Russian Academy of
Sciences, Chernogolovka, Moscow District, 142432 Russia}

%\date{}

\begin{abstract}

Superconductivity is shown to be completely destroyed in thin
mesoscopic or nanoscopic rings closed by the junction with a
noticeable interfacial pair breaking and/or a Josephson coupling,
if a ring's radius $r$ is less than the minimum radius $r_{\text{min}}$.
The quantity $r_{\text{min}}$ depends on the phase difference $\chi$ across
the junction, or on the magnetic flux that controls $\chi$ in the flux-biased 
ring. It also depends on the Josephson and interfacial effective coupling
constants, and in particular, on whether the ring is closed by $0$ or $\pi$
junction. The current-phase relation is substantially modified when the ring's
radius exceeds $r_{\text{min}}$ for some of the phase difference values,
or slightly goes beyond its maximum. The modified critical temperature $T_c$,
as well as the temperature dependent supercurrent near $T_c$ are identified here
as functions of the ring's radius and the magnetic flux.
\end{abstract}

%\pacs{74.50.+r, 74.45.+c, 74.78.Na}

\maketitle

\section{Introduction}
\label{sec: intro}

The superfluid flow in thin superconducting wires is known to result in
pair-breaking effects, which reduce the order parameter and, in
accordance with the Landau criterion, can fully destroy
superconductivity at the critical value of the superfluid velocity
$v_s$.  While the Cooper pair density diminishes with increasing $v_s$, the
supercurrent shows a nonmonotonic behavior, with its maximum
value called the depairing current $j_{\text{dp}}$. \cite{Tinkham1996}
When a thin wire forms a circular loop with radius $r$ and the absolute value
of the order parameter stays spatially constant, the order parameter-superflow
relation remains as it was in the straight wire. Specific features of the
loop topology show up, when, for example, $v_s$ is induced by the magnetic
flux $\Phi$ penetrating the loop. The flux-induced changes of the winding
number result in the oscillations of physical characteristics of the ring,
in particular, of $v_s$ and $T_c$, i.e., in the standard Little-Parks
effect \cite{LittleParks1962,*LittleParks1964,Tinkham1996}. The effect
allows a remarkably simple description within the Ginzburg-Landau (GL) 
approach, since the equilibrium order-parameter absolute value and $v_s$
are spatially constant in a cylindrically symmetric thin ring.

Constant $v_s$ is determined, along with its circulation,
by a full magnetic flux through the loop:\, $2\pi r v_s\propto\bigl(
\widetilde{\Phi}-n \bigr)$. Here $\widetilde{\Phi}=\frac{\Phi}{\Phi_0}$
is the magnetic flux in units of the superconductor flux quantum
$\Phi_0$, and $n$ is the winding number. The critical temperature $T_c$,
modified by the magnetic flux in the Little-Parks effect, can be found
taking $v_s$ in the last relation to be equal to the critical
superfluid velocity, when the order parameter vanishes and
thermodynamic potential of the ring coincides with
that of the normal metal ring. This results in the equation $r=\xi(T)
\bigl|\widetilde{\Phi}-n\bigr|$, where $\xi(T)$ is the temperature
dependent coherence length of the superconducting material. Solving the
equation with regard to the temperature, establishes the modified $T_c$
that depends on the magnetic flux, the winding number and the loop radius.
When the quantity $\bigl(\widetilde{\Phi}-n\bigr)$ is fixed, the superfluid
velocity is inversely proportional to the radius, analogously to its
dependence on the distance to the center of the Abrikosov vortex.
As a result, the minimum radius $r_{\text{min}}=\xi(T)\bigl|
\widetilde{\Phi}-n\bigr|$, at which $v_s$ takes its
critical value, exists for mesoscopic and nanoscopic uninterrupted rings.
Superconductivity is fully destroyed in the rings with radii $r\le
r_{\text{min}}$. Since the pair breaking induced by the superflow is
most pronounced at the maximum equilibrium value $\bigl|\widetilde{\Phi}
-n\bigr|=0.5$, there is no superconductor-normal metal transition, when
the magnetic flux slowly varies in the rings with $r>0.5\xi(T)$. There are
only the usual Little-Parks oscillations that occur in this case. The
transition comes about under the opposite condition, i.e., in the rings
with radii $r\le0.5\xi(T)$. It can be experimentally observed down to
quite low temperatures, when the quantum phase transition takes place
\cite{DeGennes1981,Liu2001,Oreg2010}.

This paper addresses thin superconducting loops closed by the
Josephson junction. It will be demonstrated theoretically that 
the minimum radius $r_{\text{min}}$ of thin superconducting rings 
involving the junction, unlike the case of the unbroken rings, is 
nonzero even if $v_s$ is much less than its critical value throughout 
the loop. It is the pair breaking due to the inverse 
proximity effect locally induced by the junction interface and by 
the phase difference $\chi$ across it that leads to the superconductor-normal 
metal phase transition in the rings of mesoscopic or nanoscopic size. 
The Josephson and interfacial pair breaking inevitably result in an
inhomogeneous profile of the complex order parameter, which contributes 
considerably to the gradient term in thermodynamic potential and makes a 
superconducting state energetically unfavorable in the rings with $r<
r_{\text{min}}$. 

The minimum radius is found to be a fraction of the 
temperature dependent coherence length $\xi(T)$, up to $0.5\xi(T)$. 
When the ring is closed by $0$ junction, the minimum radius, as a function
of the phase difference, is shown to have its maxima at $\chi=(2m+1)\pi$ ($m$ is an integral 
number). For $\pi$ junction, the maxima are at $\chi=2m\pi$. In both cases, 
i.e. at $\chi=m\pi$, $v_s$ vanishes together with the supercurrent all along 
the ring. It is in contrast to uninterrupted rings, where the kinetic energy
of the supercurrent becomes equal to the condensation energy and $v_s$ takes
its critical value at the superconductor-normal metal transition point. When
$r$ exceeds $r_{\text{min}}$ for some of $\chi$'s values, or goes somewhat
beyond its maximum, both the critical current and the current-phase relation
of the junction become quite sensitive to the radius value.

The phase difference across the junction will be assumed to be controlled 
by the applied magnetic flux $\Phi_e$ in the flux-biased ring. The magnetic 
field much less than the superconductor critical fields will also be supposed. 
Since the inductance effects are negligibly small near the transition point
$r=r_{\text{min}}$, where the supercurrent vanishes, the difference between the 
full magnetic flux $\Phi$ and the applied one $\Phi_e$ will be disregarded below. 
Therefore, the supercurrent hysteretic behavior due to the inductance effects 
\cite{SilverZimmerman1967,Barone1982} will not be considered. When 
the difference $r-r_{\text{min}}$ is large or moderate, the hysteretic behavior 
will be shown to appear also due to the absence of the Meissner effect in thin 
superconducting rings. This paper mainly concerns itself with the rings of smaller
sizes. Superconductivity continuously weakens in such rings and is ultimately 
destroyed, when the difference $r-r_{\text{min}}$ slowly diminishes. Once the 
difference vanishes, the superconductor-normal metal phase transition of the 
second order occurs.

Within the GL theory, the coherence length $\xi(T)$ of the superconducting material
is the only characteristic length of the problem in question, and the equation for
the minimum radius is actually formulated for the dimensionless quantity $R
=r\big/\xi(T)$. When the fixed value of $r$ slightly exceeds $r_{\text{min}}$,
the quantity $R$ can reach $R_{\text{min}}$ as the temperature goes up
making $\xi(T)$ sufficiently large. This results in the modified critical
temperature $T_c$, which depends on $r$ and $\chi$, or $\Phi$. 
%In the equation
%for $T_c$ one should take into account that the effective dimensionless
%Josephson and interface coupling constants, coming from the boundary conditions
%at the junction interface, are also temperature dependent and increase
%proportionally to $\xi(T)$. This is the consequence of the increasing influence
%of the interface on the properties of the ring as a whole, when
%the temperature draws near to $T_c$. 
The current-phase and current-flux
relations can become quite sensitive to the temperature in the narrow vicinity
of $T_c$, similarly to the presence of the radius dependence of the Josephson
current, with $r$ being quite close to $r_{\text{min}}$.

The existing temperature and magnetic flux dependence of $r_{\text{min}}$ 
near the transition make it possible to observe the phase transition by 
changing the temperature or the magnetic flux that penetrates the individual 
ring. Likely alternatives to make the effect discernible are
the junctions with interfaces made of normal metals, magnets or
other pair breaking materials, and/or the junctions involving
unconventional superconductors. The pair breaking by the phase
difference becomes noticeable for interfaces with sufficiently high
transparency. 

In the paper, Sec.~\ref{sec: eqs} addresses basic equations of the GL theory
that describe properties of thin superconducting rings closed by the Josephson junction.  
The minimum radius for such rings is obtained in Sec.~\ref{sec: rmin}.
The modified critical temperature is identified in Sec.~\ref{sec: tcmod}. 
The current-phase and current-flux relations and their dependence on the ring's radius and 
on the temperature are found in Sec.~\ref{sec: jonR}. Sec.~\ref{sec: conclusions} concludes 
the paper. Appendices~\ref{sec: sol} and~\ref{sec: mrc} present the analytical solutions of 
the equations studied. 

\section{Basic equations}
\label{sec: eqs}

Consider a superconducting circular thin ring closed by the Josephson
junction. The ring's lateral dimensions are supposed to be much less 
than $\xi(T)$ and the magnetic penetration depth. The thickness of 
the junction interface is on the order of or less than the zero temperature 
coherence length. Within the GL approach, the latter scale is considered 
to be zero. The GL free energy of the ring is represented 
as a sum of two terms:\,${\cal F}={\cal F}_{b}+{\cal F}_{\text{int}}$. 
The bulk free energy per unit area of the cross section is
\begin{multline}
{\cal F}_b\!=\!\!\!\int\limits_{0}^{2\pi r}\!\!ds\!\left[K
\left|\left(\dfrac{d}{ds}-\dfrac{2\mathtt{i}e}{\hbar c}A\right)\Psi(s)\right|^2\!\!+
\right.\\ \left.
+a\left|\Psi(s)\right|^2\!\!+\dfrac{b}{2}\left|\Psi(s)\right|^4\right]\,,
\label{F1gt}
\end{multline}
where $r$ is the ring's radius, $s=r\varphi$ is the coordinate along the 
ring's circumference and $\varphi$ is the polar angle. The coefficient in 
front of the gradient term is here denoted as $K$. The vector potential 
is taken to be cylindrically symmetric and to have only the polar component, 
i.e., the $s$-component in our notations. Such a gauge exists, for example,
for the Aharonov-Bohm flux, which is delta-localized along the ring's axis,
for the homogeneous magnetic field as well as for the one produced by
the current in a thin circular ring.

The term ${\cal F}_{\text{int}}$ is the interfacial free energy per
unit area that can be written as
\be
{\cal F}_{\text{int}}=g_{J}\left|\Psi_{0+}-\Psi_{0-}\right|^2
+g\left(\left|\Psi_{0+}\right|^2+\left|\Psi_{0-}\right|^2\right)\,.
\label{fint1gt}
\ee
The junction interface is taken at $s=0$. 

Two interface invariants
in \eqref{fint1gt} are determined both by the symmetry of the system and by 
the microscopic consideration. The latter allows one to unambiguously identify 
the two contributions to \eqref{fint1gt}, one with the Josephson coupling of 
the superconducting banks, with the coupling constant $g_J$, the other with
the interfacial pair breaking ($g>0$) that in particular takes place in the absence of the supercurrent. 
In a symmetric junction $\left|\Psi_{0+}\right|=\left|\Psi_{0-}\right|$ the first term 
in \eqref{fint1gt} takes the form $g_{J}\left|\Psi_{0}\right|^2(1-\cos\chi)$, which is known 
in decribing standard symmetric tunnel junctions. In the standard case the interfacial pair 
breaking is negligibly small (i.e., $g=0$), and a thin interface does not affect the superconductor 
at zero phase difference $\chi$ across it. Based on \eqref{F1gt} and \eqref{fint1gt}, the supercurrent 
through the Josephson junction can be described also beyond the tunneling approximation and taking 
account of the interfacial pair breaking ($g\ne0$) \cite{Barash2012,Barash2014_2,Barash2014_3}. 
The interfacial free energy \eqref{fint1gt} controls the corresponding anharmonic current-phase 
relation as well as the normalized critical current $\tilde{\jmath}_{\text{c}}=j_{\text{c}}
\big/j_{\text{dp}}$, where $j_{\text{dp}}=\bigl(8|e||a|^{3/2}K^{1/2}\bigr)\big/\bigl(3\sqrt{3}
\hbar b\bigr)$ is the depairing current deep inside the superconducting leads.

Taking the order parameter in the form $\Psi=(|a|/b)^{1/2}f(s)
e^{\mathtt{i}\phi(s)}$, one can transform the GL equation for the order parameter,
which follows from the bulk free energy \eqref{F1gt}, to the equations for $f$
\be
\dfrac{d^2f}{dx^2}-\dfrac{i^2}{f^3}+f-f^3=0
\label{gleq2}
\ee
and to the current conservation condition. Here $x=s/\xi(T)$ is the
dimensionless coordinate and $\xi(T)=(K/|a|)^{1/2}$. The 
dimensionless current density in \eqref{gleq2} is $i=\frac{2}{3\sqrt{3}}
\tilde{\jmath}=\frac{2}{3\sqrt{3}}(j\big/j_{\text{dp}})$.
                                                                                            
The quantity $f(x)$ is continuous at the interface of the
symmetric junction and has to satisfy the periodicity
condition $f(x+2\pi R)=f(x)$, where $R=r/\xi$ is the dimensionless radius. 
The boundary conditions at $x=\pm0$ that follow from \eqref{fint1gt} and 
\eqref{F1gt}, can be split into the discontinuity condition for
${df}/{dx}$ and the expression for the Josephson current via the value 
$f_{0}$ at the interface and the phase difference $\chi=\phi_{0-}-\phi_{0+}$:
\begin{align}
\biggl(&\frac{df}{dx}\!\biggr)_{\pm0}\!\!\! =\pm g_b f_0, \qquad
g_b=g_{\delta}+2g_\ell\sin^2\!\frac{\chi}2, \label{bcss99}\\
& i=-\,f^2\left(\dfrac{d\phi}{dx}+\dfrac{2\pi\xi}{\Phi_0}A\right)
=g_\ell f_{0}^2\sin\chi.
\label{bcss9}
\end{align}
Here $\Phi_0=\frac{\pi\hbar c}{|e|}$ is the magnetic flux quantum;
$g_\ell=g_J\xi(T)/K$ and $g_{\delta}=g\xi(T)/K$  are the dimensionless
Josephson and interface effective coupling constants.

The effective coupling constants play an important role in the approach
developed. Within the BCS theory, the range of variations of $g_\ell$ and
$g_\delta$ is generally quite wide \cite{Barash2012,Barash2014_2,Barash2014_3}.
Thus in dirty junctions with small or moderate transparency the quantity $g_\ell>0$
can vary from extremely small values in the tunneling limit to values that are
larger than $100$. The parameter $g_\ell$ in junctions with high transparency
can be very large. While the depairing by the interface is very weak
$g_\delta\ll 1$ in standard tunnel junctions with a conventional insulating
barrier, a pronounced depairing  $g_\delta\gg 1$ can occur in various superconductors,
including s-wave superconductors, near interfaces with normal metals and/or magnets.
In unconventional superconductors, significant depairing can also occur near the
superconductor-insulator and superconductor-vacuum interfaces.

Once $g_b\ne0$, the boundary conditions \eqref{bcss99} 
induce an inhomogeneous equilibrium profile of the order
parameter $f(x)$ in the ring, with its minimal value $f_0<1$ at $x=0$.
Since the supercurrent in thin wires is spatially constant, the modulus
and the phase of the complex order parameter vary in space
interactively, and pronounced inhomogeneities of $f(x)$ and of the
gradient of the phase along the ring take place simultaneously. To find
the current-phase relation based on \eqref{bcss9}, the self-consistent
interface value $f_0$ should be determined as a function of $R$ and $\chi$. 
The corresponding solution can be obtained based on the first integral of the 
one-dimensional GL equation \eqref{gleq2}, analogously to other problems of this type
\cite{Langer1967}. 

The analytical solutions of the GL equations that describe the order-parameter profile, 
the magnetic flux and thermodynamic potential as functions of $R$ and $\chi$, 
are obtained in Appendix~\ref{sec: sol} and used for further studies 
in the following sections. A periodic inhomogeneous 
solution for the order-parameter absolute value, with only a single minimum and a 
single maximum in the ring, is considered as being, as a rule, energetically the 
most favorable one. The minimum is induced at the junction interface $x=0$ by 
the pair breaking effects. Its derivative at $x=0$ is nonzero and discontinuous 
in accordance with the boundary conditions \eqref{bcss99}. In the equilibrium, one 
expects the maximum to be realized at the point $x=\pi R$ diametrically opposed to 
$x=0$. 

The location of the minimum at the junction interface implies an interface's pair 
breaking effect $g_b=g_\delta+2g_\ell\sin^2\frac{\chi}{2}>0$ to take place at all 
phase differences, that signifies the validity of the conditions $g_\delta>0$ and 
$-2g_\ell<g_\delta$, irrespective of the sign of the effective Josephson coupling 
constant $g_\ell$. Within this framework, the solutions obtained describe at $g_\ell<0$ 
the properties of thin superconducting rings closed by $\pi$ junction with
a pair breaking interface. Another possible solution, which corresponds to
a proximity-enhanced superconductivity near the junction interface ($g_b<0$),
at least for some of the phase differences, will not be considered in the paper,
as no evidences for thin interfaces of such type in the Josephson junctions are 
available till now.

\section{Minimum radius}
\label{sec: rmin}

Superconductivity in a thin ring closed by the junction is destroyed 
under the condition $R<R_{\text{min}}$. The dimensionless minimum radius 
$R_{\text{min}}$ appears as a consequence of the junction's destructive 
effects on the adjacent superconductivity region, i.e., of the inverse 
proximity effects underlying the boundary conditions \eqref{bcss99}. 
Solutions of \eqref{gleq2}-\eqref{bcss9} take into account the effects.

Since the second order phase transition takes place at $R=R_{\text{min}}$, the 
quantity $f(x)$ should be very small in its vicinity. A linearization 
of the GL equation is known to be the simplest way for describing 
superconducting phenomena very close to the superconductor-normal metal 
transition. One could mention in this regard, for example, the problems of 
$H_{c2}$ and $H_{c3}$ \cite{Abrikosov1957,DeGennes1963,Tinkham1996}, as well as 
of the proximity effects in the vicinity of the superconductor-normal metal
boundaries \cite{DeGennes1969,Abrikosov1988}. 

However, in the issue under consideration, the spatial dependence of both the 
absolute value of the order parameter and the gradient of its phase play an important part.
At a nonzero supercurrent density $i$, spatially constant 
in thin rings, the gauge invariant gradient of the phase (the superfluid velocity) is proportional 
to $\frac{i}{f^2}$. In agreement with this relation, $v_s(x)$ is 
a spatially dependent quantity that does not vanish at the transition point, if the phase 
difference is not a multiple of $\pi$. This substantially complicates the linearization in $f(x)$
of the GL equation \eqref{gleq2}, where the supercurrent density $i$ should be obtained 
via \eqref{bcss9} that incorporates the solution $f_0$ taken at the interface. 
The quantities $v_s(x)$, $i$ and $f_0$ are, in general, nonlocal functionals of $f(x)$
that result in the complication mentioned. Because of the second term $i^2f^{-3}$ in equation 
\eqref{gleq2}, its linearization in $f(x)$ is impossible without specifying 
the corresponding behavior of the supercurrent. Were $v_s(x)$ vanishingly small near the transition, 
the second term $i^2f^{-3}\propto v_s^2(x)f(x)$ would be negligible in \eqref{gleq2} as compared to the 
linear one. It is, however, not the case and the term, in general, can't be disregarded.

A specific case comes about, when $\chi$ is a multiple of $\pi$ and the term $i^2f^{-3}$
strictly equals to zero, together with $v_s$ and $i$. Quite close to the transition point one 
can also put $R=R_{\text{min}}$ and neglect the cubic term in \eqref{gleq2}. 
The resulting solution, for $|x|\le \pi R_{\text{min}}$, is
$f=f_d\cos(|x|-\pi R_{\text{min}})$, and the boundary conditions
\eqref{bcss99} take the form $\tan(\pi R_{\text{min}})=g_\delta$ at $\chi=2m\pi$
and $\tan(\pi R_{\text{min}})=g_\delta+2g_\ell$ at $\chi=(2m+1)\pi$. One gets from
here the quantities
\be
R_{\text{min}}(0)=\frac1{\pi}\arctan g_\delta, \quad
R_{\text{min}}(\pi)=\frac1{\pi}\arctan(g_\delta+2g_\ell),
\label{rminmax1}
\ee
which satisfy the relation $R_{\text{min}}(\pi)>R_{\text{min}}(0)$ for zero
junctions ($g_\ell>0$), while $R_{\text{min}}(\pi)<R_{\text{min}}(0)$ for
$\pi$ junctions ($g_\ell<0$). Under the condition $g_\delta\gg1$ (and/or $g_\ell
\gg1$ and $\chi\approx\pi$, in case of $0$ junctions) the minimum radius $r_{\text{min}}$
approaches its upper bound $0.5\xi(T)$. Exactly the same bound to $r_{\text{min}}$ arises
in uninterrupted mesoscopic thin rings, where $r_{\text{min}}=\xi(T)\bigl|
\widetilde{\Phi}-n\bigr|$ and the maximum equilibrium value of $\bigl|\widetilde{\Phi}-
n\bigr|$ is $0.5$. It is also noted that the quantity $r_{\text{min}}(0)$ is related to
the minimum length $L_{\text{min}}$ of a thin straight superconducting wire or a thin film
symmetrically sandwiched between identical pair breaking walls:
$L_{\text{min}}=2\pi r_{\text{min}}(0)=2\xi(T)\arctan g_\delta$ \cite{Ginzburg1958,ROZaitsev1965,Ginzburg1993}.

Moving over to a more general description, one could proceed by taking into
account the nonlinear second term, but neglecting from the very beginning the
cubic term in \eqref{gleq2}. The corresponding simplified solution is expressed
via the inverse trigonometric functions. Alternatively, one can find the minimum 
radius by introducing the corresponding simplifications in the exact solution of 
the GL equations \eqref{gleq2}-\eqref{bcss9} obtained in Appendix~\ref{sec: sol}. 
The former approach works well just at the transition point and it does not apply 
to the vicinity of the transition, which will be studied in Sec.~\ref{sec: jonR}. 
Since the exact solution is in any case required for this paper, it is also used 
for the present purpose in Appendix~\ref{sec: mrc}, where the dimensionless minimum 
radius, obtained as a function of the phase difference as well as of the Josephson and 
interface effective coupling constants, is shown to be described by the following expression
\footnote{Expressions \eqref{ttt3} and \eqref{sol8} are
slightly simplified, if the function $\arccot x$ is used instead of
$\arccos x$, while their principal values are defined within the same
range $(0,\pi)$. Here the function $\arccos x$ is chosen, since, for
instance, in \cite{Wolfram2003} the range $(0,\pi)$ is used only for
$\arccos x$, while for $\arccot x$ the range $(-\pi/2,\pi/2)$ has been
taken.}
\be
R_{\text{min}}\!=\!\dfrac{1}{2\pi}
\arccos\left(\!\dfrac{1-g^2_{\ell}\sin^2\!\chi-g_b^2}{\sqrt{\smash[g]{
\left(1-g^2_{\ell}\sin^2\!\chi-g_b^2\right)^2+4g_b^2}}}\!\right).
\label{ttt3}
\ee
As defined in \eqref{bcss99}, $g_b=g_{\delta}+2g_\ell\sin^2\!\frac{\chi}2$.
Simple analytical expression \eqref{ttt3}, describing the
minimum radius $R_{\text{min}}$ as a function of $\chi$, $g_\ell$
and $g_\delta$, is one of the prime results of the paper.

The continuous character of the phase transition is confirmed at sufficiently 
small $R-R_{\text{min}}$ by the order-parameter behavior\,$f_0^2\propto
(R-R_{\text{min}})$. The free energy linearly vanishes with $(R-R_{\text{min}})$ 
near the transition point
\be
\widetilde{\cal F}\approx-\,\dfrac{4\pi}{3}\bigl(R-R_{\text{min}}\bigr).
\label{ttt4}
\ee
It is the result of a strong competition between the interfacial and bulk 
contributions, that takes place in mesoscopic and nanoscopic rings in the presence of 
the inverse proximity effects.

As follows from \eqref{ttt3} for the rings closed by $0$ junction, the phase
dependence of $R_{\text{min}}$ becomes noticeable, when the interface pair breaking
is not too strong $g_\delta\alt1$. The larger the strength
of the Josephson coupling $g_\ell>0$, the more pronounced modulation of
$R_{\text{min}}$ with the phase difference that takes place in a wide ($g_\ell^2 
\alt1$) or narrow ($g_\ell^2\gg 1$) vicinity of $\chi=2m\pi$. The difference between
$R_{\text{min}}(0)$ and $R_{\text{min}}(\pi)$, and, therefore,
the phase dependence of $R_{\text{min}}$ as a whole, become negligibly small, when
$g_\delta\gg1$ (see also \eqref{rminmax1}). Since $R_{\text{min}}$
increases with $g_\ell$ for $0$ junctions, its minima stay at $\chi=2\pi m$
and do not depend on $g_\ell$. Its maxima are at $\chi=(2m+1)\pi$ and depend on both
coupling constants $g_\ell$ and $g_\delta$.

As $g_\ell<0$ in $\pi$ junctions and $g_b>0$ is assumed, a finite Josephson coupling
term $2g_\ell\sin^2\frac{\chi}{2}$ reduces the strength of the pair breaking by the
junction interface characterized by the parameter $g_b$. Correspondingly,
the quantity $R_{\text{min}}$ decreases with increasing strength of the Josephson coupling
$|g_\ell|$. Its minima occur at $\chi=(2m+1)\pi$ and depend on both values
$g_\ell$ and $g_\delta$, while the maxima are at $\chi=2m\pi$ and depend solely on
$g_\delta$. Since the supercurrent \eqref{bcss9} and the superfluid velocity vanish
at $\chi=m\pi$, the results regarding the extrema of $R_{\text{min}}$ agree with
\eqref{rminmax1} found for $\chi=0,\pi$ within the linearized description. In accordance
with \eqref{ttt4}, free energy has its minimum at $\chi=2m\pi$ for $0$ junction, 
and at $\chi=(2m+1)\pi$ for $\pi$ junction.

\begin{figure}[thb!]
\begin{center}
\begin{minipage}{.49\columnwidth}
\includegraphics*[width=.99\columnwidth,clip=true]{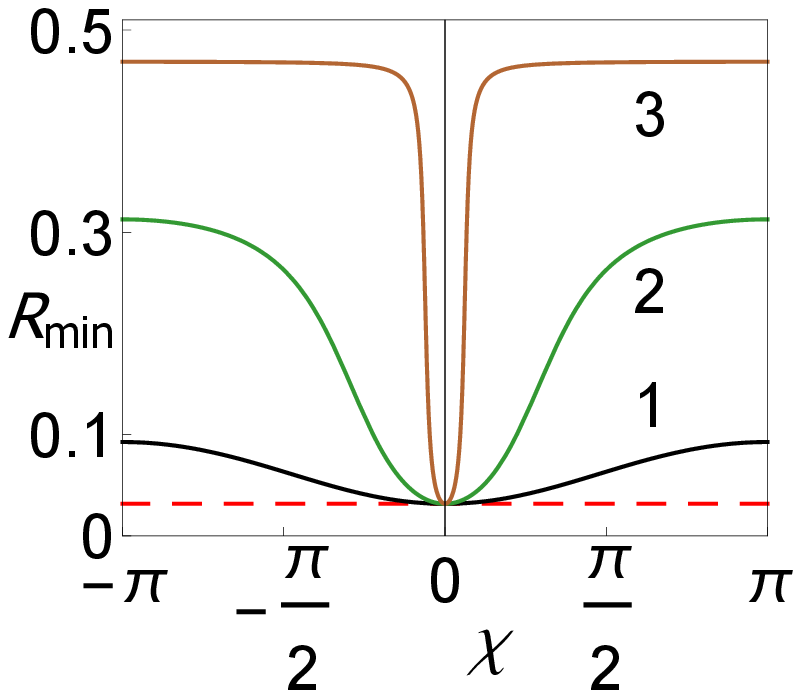}
\end{minipage}
\begin{minipage}{.49\columnwidth}
\includegraphics*[width=.97\columnwidth,clip=true]{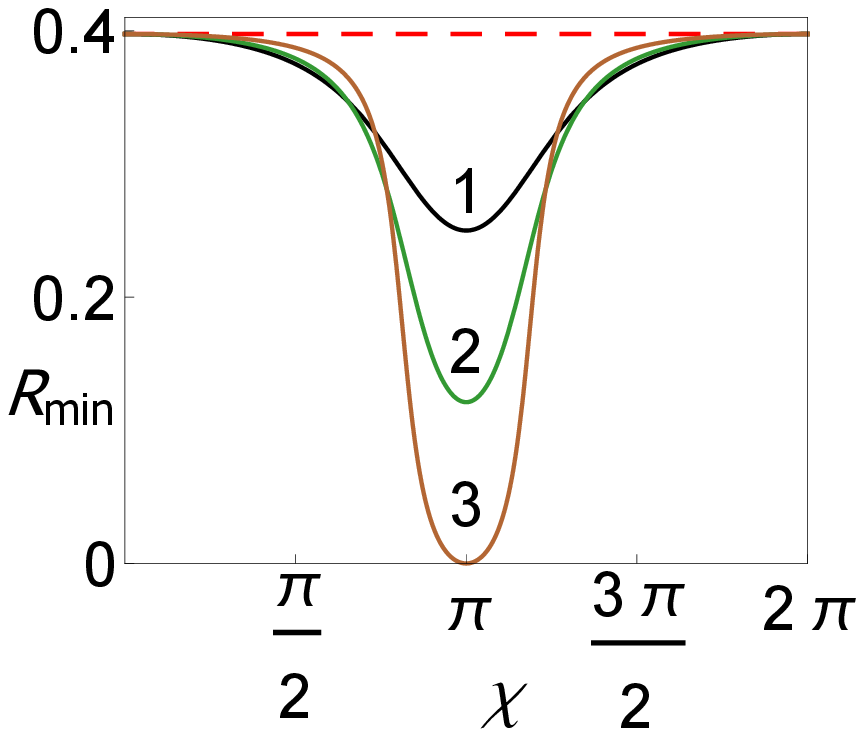}
\end{minipage}
\end{center}
\caption{$R_{\text{min}}(\chi)$ for the rings closed by $0$ junction (left panel) and
$\pi$ junction (right panel). {\it Left panel}: $g_\delta=0.1$ and  \,\,
(1)\, $g_{\ell}=0.1$\,\,
(2)\, $g_{\ell}=0.7$\,\,
(3)\, $g_{\ell}=5$,\,\, and $g_\ell=0$ - dashed curve.
{\it Right panel}:
$g_\delta=3$ and \,\,
(1)\, $g_{\ell}=-1$\,\,
(2)\, $g_{\ell}=-1.3$\,\,
(3)\, $g_{\ell}=-1.5$,\,\, and $g_\ell=0$ - dashed curve.
}
\label{fig:rminchi}
\end{figure}
The minimum radius $R_{\text{min}}$ as a function of the phase
difference, for the thin rings closed by $0$ junctions and $\pi$ junctions,
is shown in the left and right panels of Fig. \ref{fig:rminchi} respectively.
The minimum radius has been found above by comparing the superconducting
rings with various radii at fixed $\chi$. When the Josephson coupling 
induces a noticeable phase dependence of $R_{\text{min}}$, the critical value 
$R=R_{\text{min}}(\chi)$ is not necessarily to be the minimum radius of the
superconducting rings with $\chi'\ne\chi$.

In the flux-biased rings the minimum radius actually depends on $\widetilde{\Phi}$, 
rather than on $\chi$. The relationship between the quantities $\chi$ and
$\widetilde{\Phi}$ can be obtained by integrating the first expression for the 
current in \eqref{bcss9} along the ring. Since $\oint_{{\cal C}}\mathbf{A}\cdot 
\xi dx=\Phi$ is the full magnetic flux penetrating the ring, one gets 
\be
\chi+2\pi\left(\widetilde{\Phi}-n\right)=-\,2i\int\limits_{0}^{\pi R}\dfrac{dx}{f^2(x)}.
\label{sol5}
\ee
Here $n$ is the winding number and the relation $f(x)=f(-x)$ has been used.

The integration on the right hand side in \eqref{sol5}, taken with the 
solution for the order parameter,  results in \eqref{sol6}, which is valid 
at arbitrary ring's size. The expression \eqref{sol6} is substantially simplified 
at $R=R_{\text{min}}(\chi)$, as shown in Appendix~\ref{sec: mrc} \cite{Note1}:
\begin{multline}
\widetilde{\Phi}-n=-\dfrac1{2\pi}\Biggl(\chi+\sgn(g_\ell\sin\chi)\times\\ 
\arccos\frac{1+g_b^2-g^2_{\ell}\sin^2\chi}{\sqrt{\Bigl(1+g_b^2-g^2_{\ell}
\sin^2\chi\Bigr)^2+4g_b^2g_{\ell}^2\sin^2\chi}}\Biggr).
\label{sol8}
\end{multline}

The relationships between the magnetic flux and the phase difference, 
obtained in the paper, can strongly deviate from the linear dependence
$\chi=-2\pi(\widetilde{\Phi}-n)$ taking place in thick rings closed by the
Josephson junction, in the absence of noticeable inductance effects
\cite{Barone1982}. Due to the Meissner effect, the supercurrent vanishes
along with $v_s$ in the depth of a thick ring, while the order
parameter remains finite. On the contrary, in thin rings the supercurrent
vanishes at $R=R_{\text{min}}$ together with the order parameter, while
$v_s$, in general, stays finite. As can be seen in \eqref{sol5}, where the
right hand side represents circulation of $v_s$, the superfluid velocity
is responsible for the nonlinear character of the relationship \eqref{sol8}.
The integral $\int{dx}\big/{f^2(x)}$ on the right hand side
of \eqref{sol5} diverges at the transition point. However, the integral
multiplied by the supercurrent $i=g_\ell f^2_0\sin\chi$, remains finite
even in the limit $f(x)\to0$, in accordance with the relation $f_0\le f(x)$. 
Both sides in \eqref{sol5} take zero values, along with $v_s$,
when $\chi$ is a multiple of $\pi$. In this particular case the relation
\eqref{sol8} acquires the familiar linear character and $|\widetilde{\Phi}|$
takes an integer or a half-integer number. For example, for $|\chi|=\pi$
(and $n=0$) one gets $|\widetilde{\Phi}|=\frac12$. Therefore, the free
energy \eqref{ttt4} has its minimum at $\widetilde{\Phi}=0$ in the case of
$0$ junction, and at $|\widetilde{\Phi}|=\frac12$ in the case of $\pi$ junction.
This agrees with the emergence of the spontaneous magnetic flux penetrating the
ring closed by the $\pi$ junction \cite{Bulaevskii1977}.

\begin{figure}[thb!]
\includegraphics[width=0.6\columnwidth,clip=true]{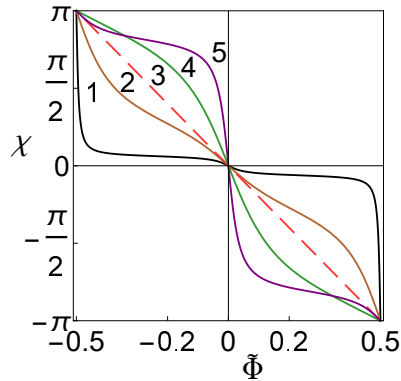}
\caption{$\chi(\widetilde{\Phi})$
in the rings of minimum radius. {\it $0$ junctions}: $g_\delta=0.1$ and \,\,
(1)\, $g_{\ell}=5$\,\,
(2)\, $g_{\ell}=0.7$\,\,
(3)\, $g_{\ell}\ll1$.\,\, {\it $\pi$ junctions}: $g_\delta=3$ and \,\,
(3)\, $|g_{\ell}|\ll1$\,\,
(4)\, $g_{\ell}=-1$\,\,
(5)\, $g_{\ell}=-1.5$.
}
\label{fig:chivsPhiatRmin}
\end{figure}

The magnetic flux dependence $\chi(\widetilde{\Phi})$ in the minimum radius rings 
is shown in Fig.~\ref{fig:chivsPhiatRmin} for $|\widetilde{\Phi}|<0.5$, $|\chi|<\pi$.
If the Josephson coupling is sufficiently weak $|g_\ell|\ll1$, then the phase difference 
$\chi$ depends almost linearly on the magnetic flux $\widetilde{\Phi}$, as follows from 
\eqref{sol8} and seen in the dashed curve in Fig.~\ref{fig:chivsPhiatRmin}. At the same time,
curves 1 and 5 in Fig.~\ref{fig:chivsPhiatRmin}, which correspond to $0$ and 
$\pi$ junctions with comparatively large Josephson coupling strengths $|g_\ell|$,
demonstrate opposite signs of pronounced deviations from the linear behavior.
Curves 1 and 5 show that the phase difference varies weakly in the wide region of $\widetilde{\Phi}$,
while it undergoes abrupt changes in the narrow vicinities of the half-integer (integer) values
of $\widetilde{\Phi}$ in $0$ junctions ($\pi$ junctions).

\begin{figure}[thb!]
\begin{center}
\begin{minipage}{.49\columnwidth}
\includegraphics[width=0.99\columnwidth,clip=true]{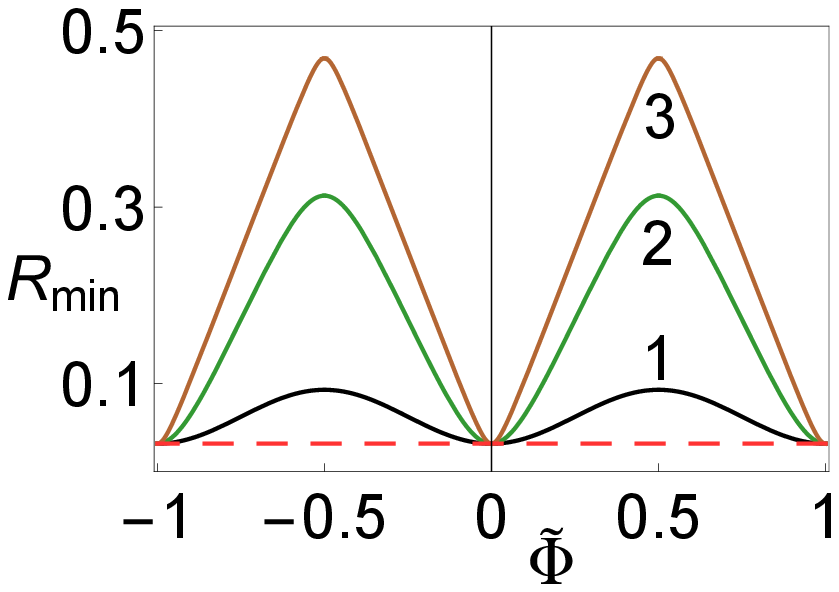}
\end{minipage}
\begin{minipage}{.49\columnwidth}
\includegraphics*[width=.99\columnwidth,clip=true]{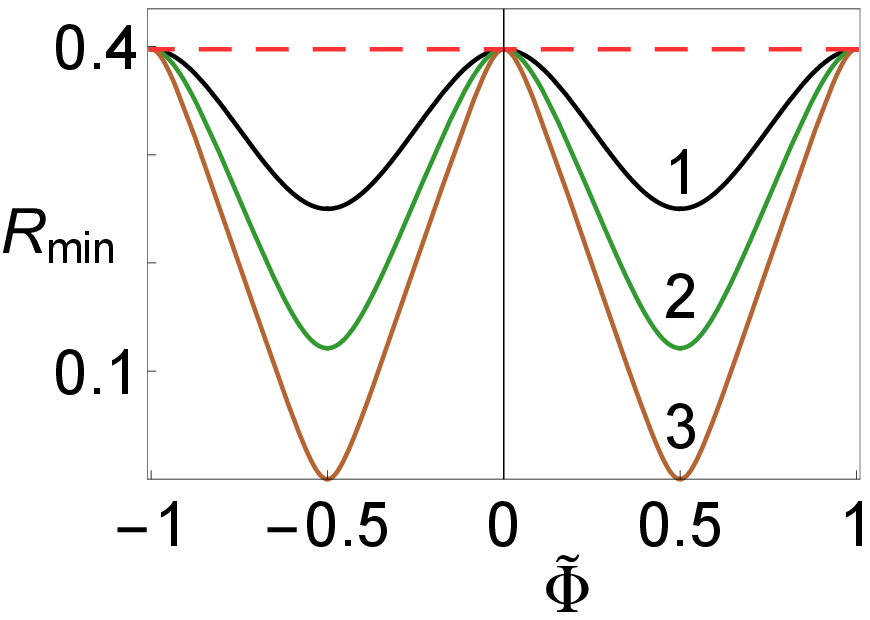}
\end{minipage}
\end{center}
\caption{$R_{\text{min}}(\widetilde{\Phi})$ for the rings with $0$ junctions
(left panel) and $\pi$ junctions (right panel) under the same set of
parameters as in the corresponding panels in Fig. \ref{fig:rminchi}.
}
\label{fig:rminphi}
\end{figure}
A combined consideration of \eqref{ttt3} and \eqref{sol8}
results in the periodic dependence of the minimum radius on the magnetic flux
$R_{\text{min}}(\widetilde{\Phi})$, which is depicted in
Fig. \ref{fig:rminphi}. The stronger the Josephson coupling strength is,
the more intense the modulation of $R_{\text{min}}$ with
the magnetic flux. In the case of $0$ junction ($\pi$-junction), 
the maxima of $R_{\text{min}}$ and of the pair breaking parameter $g_b$
occur at the magnetic flux half-integer (integer) values.

\section{Modified critical temperature}
\label{sec: tcmod}

Since temperature has been incorporated in the definitions of a
number of dimensionless quantities, it implicitly enters all the results
obtained above. Thus the temperature dependent coherence
length $\xi(T)$, as the characteristic length of the inverse proximity effects,
is included in the dimensionless radius $R(T)=r\big/\xi(T)$. When $R(T)$
is taken at fixed $r$, it decreases with diminishing $\tau$, and, at least
formally, vanishes at $\tau=0$. Here $\tau=1-(T/T_{c0})$, and
$T_{c0}$ is an unperturbed critical temperature of the superconducting material.

On the other hand, the right hand side of \eqref{ttt3}, i.e., $R_{\text{min}}$,
depends on $\tau$ via the temperature dependent effective coupling constants
$g_\ell=(g_J \xi(T))\big/K$ and $g_\delta=(g \xi(T))\big/K$. They increase with 
$\xi(T)$, when the temperature draws near to $T_c$, as a consequence of an 
increasing influence of the junction interface and, therefore, the boundary 
conditions \eqref{bcss99}, on the ring's properties as a whole. When $\tau$ goes 
down, the dimensionless radius decreases down to $0$ at $\tau=0$, while the quantity
$R_{\text{min}}$ increases up to $0.5$, as seen in \eqref{rminmax1} and \eqref{ttt3}.

Therefore, if the ring's radius $R$ initially exceeds $R_{\text{min}}$ and
temperature goes up, making $\xi(T)$, $g_\ell$ and $g_\delta$ sufficiently large
and $\tau$ small, the quantities $R$ and $R_{\text{min}}$ will inevitably become
equal to each other. The temperature $T_c$, at which the equality takes place, is
the modified critical temperature of the superconducting transition
in the ring closed by the junction. In other words, the equality
$R=R_{\text{min}}$, with $R_{\text{min}}$ given in \eqref{ttt3}, can also be
considered as
the equation for the proximity-modified critical temperature $T_c<T_{c0}$ that
depends on the ring's radius and the phase difference. The critical temperature
shift $\Delta T_c=T_c-T_{c0}$ is discernible within the mean field theory, when
exceeding the fluctuation region near $T_{c0}$. This is the case for
mesoscopic or nanoscopic rings, when the ring's radius slightly
differs from the minimum one. For a sufficiently large $R$ the
shift is negligibly small.

In order to reveal the temperature dependence in the equation
$R=R_{\text{min}}$, let us introduce an auxiliary ``low-temperature''
GL coherence length $\xi_0^{\text{GL}}=\xi(T)\sqrt{\tau}$\,
\footnote{For a comparison of $\xi_0^{\text{GL}}$ with the 
Bardeen-Cooper-Schrieffer coherence length $\xi_0$ see, for example,
Ref. \onlinecite{Golubov2001}} and move on to a rescaled dimensionless
radius $\widetilde{R}=r\big/\xi_0^{\text{GL}}=R\big/\sqrt{\tau}$.
The effective ``low-temperature'' coupling constants of the GL theory
are defined as $g_{\ell0}=(g_J \xi_0^{\text{GL}})\big/K$, $g_{\delta0}=
(g\xi_0^{\text{GL}})\big/K$. With these definitions, one gets from
\eqref{ttt3} the following relationship
\be
\widetilde{R}\!=\dfrac{1}{2\pi\sqrt{\tau}}
\arccos\dfrac{\tau-g^2_{\ell0}\sin^2\!\chi-g^2_{b0}}{\sqrt{\smash[bg]{
\left(\tau-g^2_{\ell0}\sin^2\!\chi-g^2_{b0}\right)^2+4g^2_{b0}\tau}}}\,,
\label{sol9}
\ee
which can be used either as the expression for the temperature dependent
minimum radius $\widetilde{R}_{\text{min}}\bigl(\frac{T}{T_{c0}},\chi\bigr)$, or as the equation
for the critical temperature shift $\tau_c(\widetilde{R},\chi)=1-
\frac{T_c(\widetilde{R},\chi)}{T_{c0}}=-\frac{\Delta T_c(\widetilde{R},\chi)}{T_{c0}}$ in the ring
with radius $\widetilde{R}$.

\begin{figure}[thb!]
\begin{center}
\begin{minipage}{.49\columnwidth}
\includegraphics*[width=.95\columnwidth,clip=true]{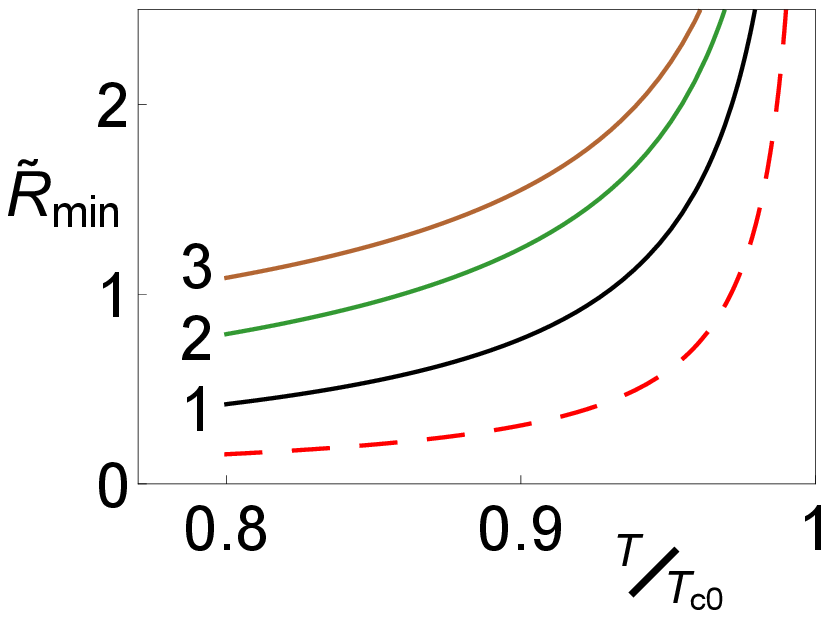}
\end{minipage}
\begin{minipage}{.49\columnwidth}
\includegraphics*[width=.95\columnwidth,clip=true]{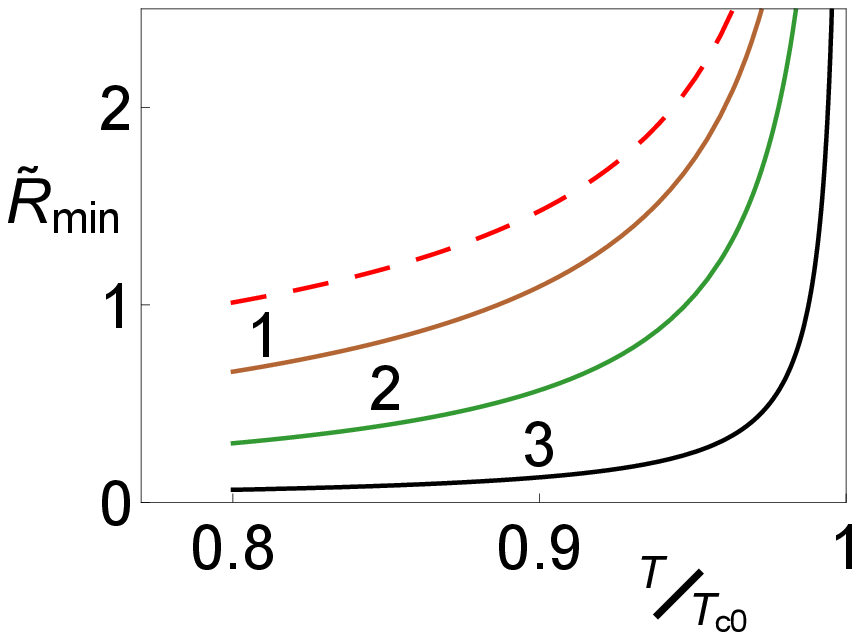}
\end{minipage}
\end{center}
\caption{$\widetilde{R}_{\text{min}}(\frac{T}{T_{c0}})$ at $\chi=0,\pi$ for rings with
$0$ junction (left panel) and $\pi$ junction (right panel).
{\it Left panel}:\,\,$g_{\delta0}=0.1$,\,\,$\chi=\pi$ and
(1)\, $g_{\ell0}=0.1$\,\,
(2)\, $g_{\ell0}=0.4$\,\,
(3)\, $g_{\ell0}=5$;\,\,$\chi=0$ - dashed curve.\,
{\it Right panel}:\,\,$g_{\delta0}=3$,\,\,$\chi=\pi$ and
(1)\, $g_{\ell0}=-1.2$\,\,
(2)\, $g_{\ell0}=-1.4$\,\,
(3)\, $g_{\ell0}=-1.48$;\,\,$\chi=0$ - dashed curve.
}
\label{fig:rmintau}
\end{figure}

At a given temperature, the extrema of $\widetilde{R}_{\text{min}}\bigr(\frac{T}{T_{c0}},\chi\bigr)$
take the form
\begin{align}
&\widetilde{R}_{\text{min}}\Bigl(\frac{T}{T_{c0}},\pi\Bigr)=\dfrac{1}{2\pi\sqrt{\tau}}
\arccos\dfrac{\tau-(g_{\delta0}+2g_{\ell0})^2}{\tau+(g_{\delta0}+2g_{\ell0})^2}\,,\label{sol9a}\\
&\widetilde{R}_{\text{min}}\Bigl(\frac{T}{T_{c0}}, 0\Bigr)=\dfrac{1}{2\pi\sqrt{\tau}}
\arccos\dfrac{\tau-g_{\delta0}^2}{\tau+g_{\delta0}^2}\,.
\label{sol9b}
\end{align}
The temperature dependence of the extrema of $\widetilde{R}_{\text{min}}(\chi)$ near $T_c$
is depicted in Fig.~\ref{fig:rmintau} for $0$ junctions (the left panel) and $\pi$ junctions
(the right panel).
For $0$ junctions, the temperature dependent minimal value
$\widetilde{R}_{\text{min}}\bigr(\frac{T}{T_{c0}},0\bigr)$ is
independent of $g_{\ell0}$, being the same for $g_{\delta0}=0.1$ and various values of $g_{\ell0}$
considered in the left panel in Fig.~\ref{fig:rmintau} (the dashed curve).
The solid curves in the Fig.~\ref{fig:rmintau} left panel demonstrate the
temperature dependent maximum $\widetilde{R}_{\text{min}}\bigr(\frac{T}{T_{c0}},\pi\bigr)$
at various values of $g_{\ell0}$ for $g_{\delta0}=0.1$. For $\pi$ junctions, the temperature
dependent maximum $\widetilde{R}_{\text{min}}\bigr(\frac{T}{T_{c0}},0\bigr)$ is
independent of $g_{\ell0}$, being the same for $g_{\delta0}=3$ and various values of $g_{\ell0}$
considered in the right panel of Fig.~\ref{fig:rmintau} (the dashed curve).
The solid curves in the right panel of Fig.~\ref{fig:rmintau} demonstrate the
temperature dependent minimum $\widetilde{R}_{\text{min}}\bigr(\frac{T}{T_{c0}},\pi\bigr)$,
taken at various negative values of $g_{\ell0}$ for $g_{\delta0}=3$.

As seen in \eqref{sol9}, the upper bound on the shift of the critical
temperature is $\tau_c\le 1/(4\widetilde{R}^2)$. While $\xi(T)$ diverges
at $T=T_{c0}$, it stays finite at $T=T_c<T_{c0}$. Taking jointly the upper
bound obtained and the simplest condition $\tau_c\ll 1$ for the GL theory to be
applied, one gets $4\widetilde{R}^2\gg1$, which also agrees with the
applicability domain of the GL theory.

\begin{figure}[thb!]
\begin{center}
\begin{minipage}{.49\columnwidth}
\includegraphics*[width=\columnwidth,clip=true]{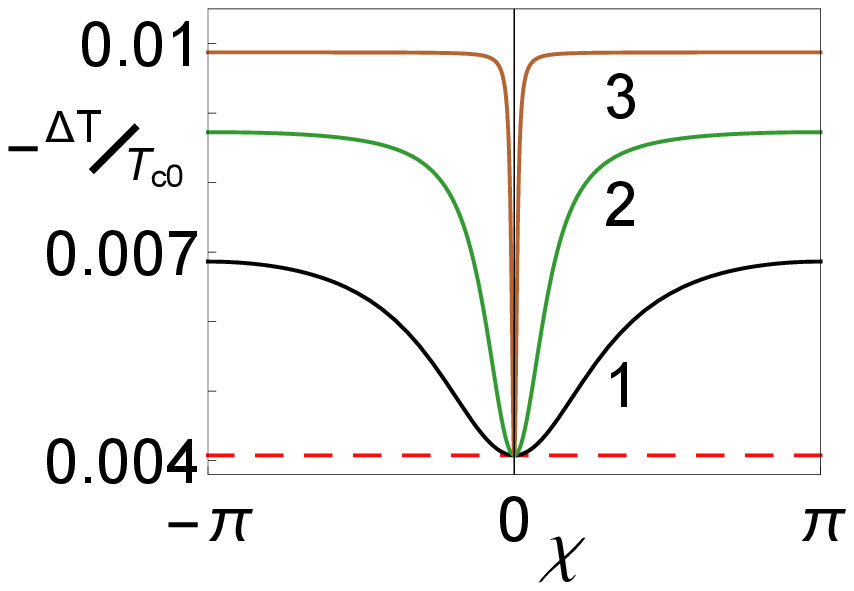}
\end{minipage}
\begin{minipage}{.49\columnwidth}
\includegraphics*[width=\columnwidth,clip=true]{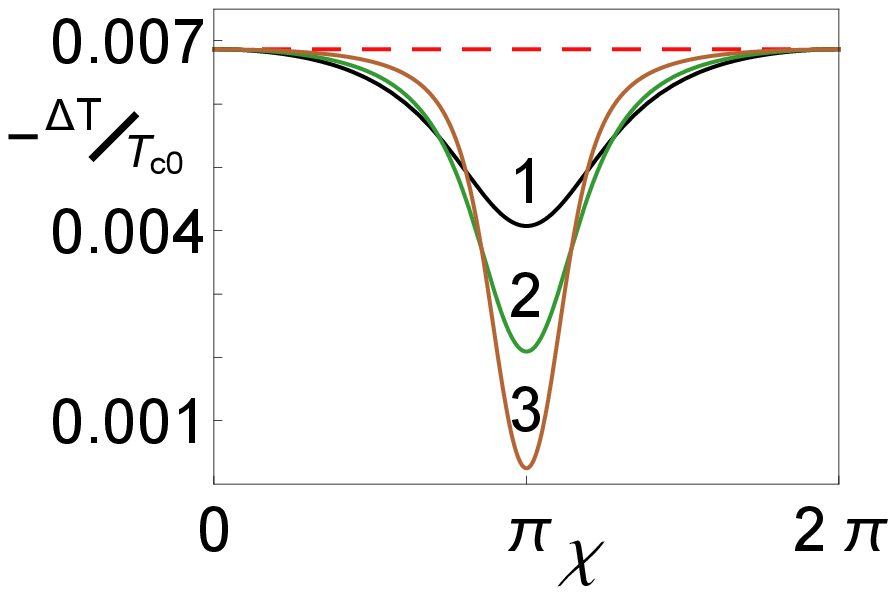}
\end{minipage}
\end{center}
\caption{$-\, \frac{\Delta T_c}{T_{c0}}$ as a function of $\chi$
in rings with $0$ junction (left panel) and $\pi$ junction (right panel); $\widetilde{R}=5$. 
{\it Left panel}:~$g_{\delta0}=0.1$ and \,\,
(1)\, $g_{\ell0}=0.1$\,\,
(2)\, $g_{\ell0}=0.4$\,\,
(3)\, $g_{\ell0}=5$;\,\, $g_{\ell0}=0$ - dashed line.
{\it Right panel}:~$g_{\delta0}=0.3$ and \,\,
(1)\, $g_{\ell0}=-0.1$\,\,
(2)\, $g_{\ell0}=-0.13$\,\,
(3)\, $g_{\ell0}=-0.148$;\,\, $g_{\ell0}=0$ - dashed line.
}
\label{fig:taucvschi}
\end{figure}

One notes, for example, that under the particular conditions
$\tau_c\ll \min\bigl(1,g_{b0}^2+g^2_{\ell0}\sin^2\chi\bigr)$, the solution
of equation \eqref{sol9} takes the form
\be
\tau_c(\widetilde{R},\chi)=-\dfrac{\Delta T_c}{T_{c0}}\approx\dfrac{\pi^2}{4\left(\pi\widetilde{R}+
\dfrac{g_{b0}(\chi)}{g_{b0}^2+g^2_{\ell0}\sin^2\chi}\right)^2}.
\label{sol10}
\ee
The critical temperature shift $\tau_{c}(\widetilde{R},\chi)$ as a
function of the phase difference is shown in Fig.~\ref{fig:taucvschi} for
the rings with $\widetilde{R}=5$, closed by $0$ junction (left panel) and
$\pi$ junction (right panel). For $0$ junction ($\pi$ junction) 
the shift takes its maxima at $\chi=(2m+1)\pi$ ($\chi=2m\pi$) and minima at
$\chi=2m\pi$ ($\chi=(2m+1)\pi$).

\begin{figure}[thb!]
\begin{center}
\begin{minipage}{.49\columnwidth}
\includegraphics*[width=\columnwidth,clip=true]{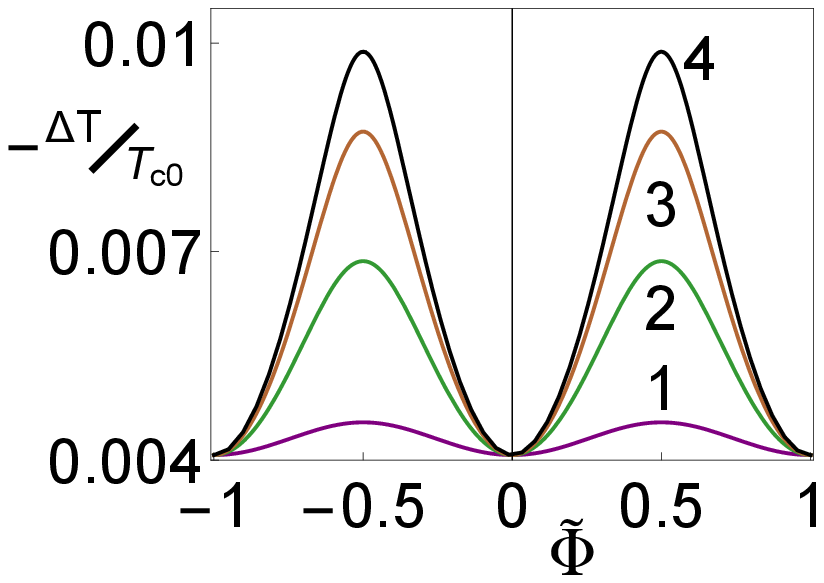}
\end{minipage}
\begin{minipage}{.49\columnwidth}
\includegraphics*[width=\columnwidth,clip=true]{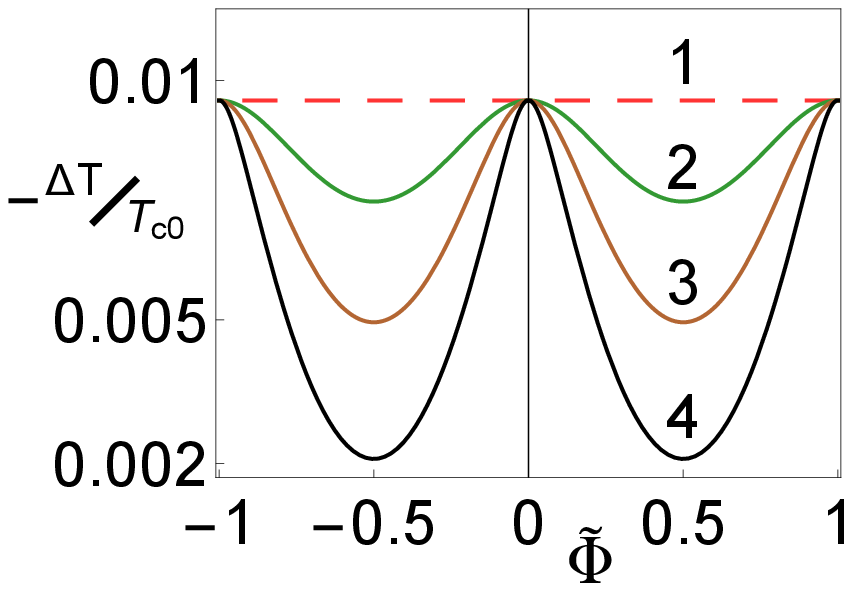}
\end{minipage}
\end{center}
\caption{$-\, \frac{\Delta T_c}{T_{c0}}$ as a periodic function of $\widetilde{\Phi}$ in
a ring with radius $\widetilde{R}=5$, closed by $0$ junction (left panel), and $\pi$ junction 
(right panel).
{\it Left panel}: $g_{\delta0}=0.1$ and \,\,
(1)\, $g_{\ell0}=0.01$\,\,
(2)\, $g_{\ell0}=0.1$\,\,
(3)\, $g_{\ell0}=0.4$\,\,
(4)\, $g_{\ell0}=5$.
{\it Right panel}: $g_{\delta0}=3$ and  \,\,
(1)\, $g_{\ell0}=0$\,\,
(2)\, $g_{\ell0}=-1.3$\,\,
(3)\, $g_{\ell0}=-1.43$\,\,
(4)\, $g_{\ell0}=-1.48$.
}
\label{fig:taucphic}
\end{figure}

The dependence of the magnetic flux $\Phi_c(\widetilde{R},\chi)$, taken at the
modified critical temperature, can be obtained by extracting the
temperature dependence in equation \eqref{sol8} and substituting
for $\tau$ the solution of \eqref{sol9} $\tau_{c}(\widetilde{R},\chi)$.
This allows one to get the relative shift of the critical
temperature $-\, \frac{\Delta T_c}{T_{c0}}$ as a periodic
function of the magnetic flux $\widetilde{\Phi}$, shown in 
Fig.~ \ref{fig:taucphic} for $0$ junction (the left panel)
and $\pi$ junction (the right panel). As expected, the
shift of $T_c$ obtained is comparatively small but can lie
beyond the fluctuation region near $T_c$, in a wide range
of $g_{\delta0}$ and $g_{\ell0}$ variations.

\section{Radius-dependent Josephson current}
\label{sec: jonR}

A noticeable dependence of the Josephson current on the ring's radius
appears, when $r$ becomes close to the minimum radius. Since
$r_{\text{min}}$ depends on the phase difference, not only the
critical current, but also the current-phase relation of the junction
can be strongly modified, when $r$ either exceeds the minimum radius for 
some of phase differences, or goes slightly over its maximum. With 
increasing $r$ at $r\agt\xi(T)$, the current-phase relation quickly 
approaches the one describing the junctions included in asymptotically 
large rings, or in straight long superconducting leads.

\begin{figure}[thb!]
\includegraphics[width=0.6\columnwidth,clip=true]{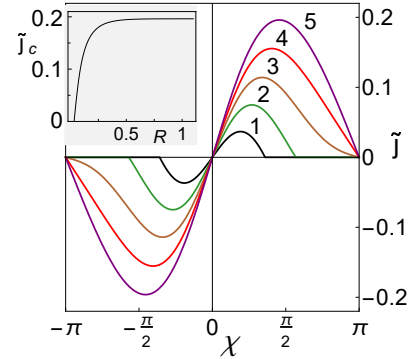}
\caption{$\tilde{\jmath}(\chi)$  for $0$ junction:\,($g_\ell=g_\delta=0.1$)
included in the rings with various $R$:\,\,
(1)\, $R=0.05$\,\,
(2)\, $R=0.07$\,\,
(3)\, $R=0.1$\,\,
(4)\, $R=0.16$\,\,
(5)\, $R\gg 1$.
{\it Inset:} The critical current $\tilde{\jmath}_c$ as a function of $R$.}
\label{fig:0jvsvarphiandjcvsR}
\end{figure}
The normalized circulating supercurrent $\tilde{\jmath}=j/j_{\text{dp}}$ 
is depicted as a function of the phase difference in the main panel of
Fig.\,\ref{fig:0jvsvarphiandjcvsR} for various ring's radii and for $0$ junction
with $g_\ell=g_\delta=0.1$. The numerical results have been obtained by carrying 
out the evaluation of the supercurrent \eqref{bcss9} with the exact 
self-consistent formulas of Appendix~\ref{sec: sol}. For the set of parameters 
chosen, one gets from \eqref{ttt3} $r_{\text{min}}(0)\approx 0.0317\xi(T)$ and
$r_{\text{min}}(\pi)\approx 0.093\xi(T)$. Curves 1-3 in
Fig.\,\ref{fig:0jvsvarphiandjcvsR} correspond to the condition
$r_{\text{min}}(0)<r\alt r_{\text{min}}(\pi)$, while curve 5
describes the current-phase relation of the same junction included in
a large ring. The dependence of the critical current on the ring's radius
is shown for the same set of parameters in the inset in
Fig.\,\ref{fig:0jvsvarphiandjcvsR}. The critical current vanishes at
$r=r_{\text{min}}(0)$, while at $r\ge\xi(T)$ its value 
is quite close to the asymptotic one.

\begin{figure}[thb!]
\includegraphics[width=0.6\columnwidth,clip=true]{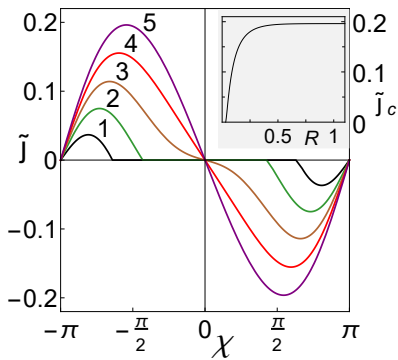}
\caption{$\tilde{\jmath}(\chi)$ for $\pi$ junction \,($g_\ell=-0.1$
and $g_\delta=0.3$) included in the rings with various $R$:\,\,
(1)\, $R=0.05$\,\,
(2)\, $R=0.07$\,\,
(3)\, $R=0.1$\,\,
(4)\, $R=0.16$\,\,
(5)\, $R\gg 1$.
{\it Inset:} The critical current $\tilde{\jmath}_c$ as a function of $R$.}
\label{fig:pijvsvarphiandjcvsR}
\end{figure}

The current-phase relations of $\pi$ junction
with $g_\ell=-0.1$ and $g_\delta=0.3$, which closes the rings with the same
set of radii, are shown in the main panel of Fig.\,\ref{fig:pijvsvarphiandjcvsR}.
For curves 1-3 in Fig.\,\ref{fig:pijvsvarphiandjcvsR}, $\pi$ junction destroys
superconductivity in the rings in a vicinity of $\chi=2m\pi$, while the supercurrent
still survives at phase differences closer to $\chi=(2m+1)\pi$. The dependence of the critical
current on the ring's radius is similar to the case of $0$ junction.
For $g_\ell=-0.1$ and $g_\delta=3$, the critical current vanishes at $r= 
r_{\text{min}}(\pi)\approx 0.0317\xi(T)$,
while at $r\ge\xi(T)$ its value is quite close to the asymptotic one.

The magnetic flux dependence of the Josephson current can be found,
in the case of $0$ junction, combining the phase dependence 
of the supercurrent shown in Fig.\,\ref{fig:0jvsvarphiandjcvsR}
with such a dependence of the magnetic flux given by \eqref{sol6}.
Unlike the current-phase relation, the current-magnetic flux
relation in thin superconducting rings substantially depends on the
radius even at large $R$. In the absence of the Meissner 
screening, the circulation of $v_s$ described on the right hand side 
of \eqref{sol5}, enters and can considerably change the relation 
between $\Phi$ and $\chi$. As was shown in Sec. \ref{sec: rmin}, the 
circulation of $v_s$ noticeably modifies the $\Phi-\chi$ relation even 
at small $r$, i.e., quite close to the transition point $r=r_{\text{min}}$, 
where the superfluid velocity does not vanish. At sufficiently large $r$, which 
enters the upper limit of the integration in \eqref{sol5}, the ``$v_s$-term'' 
in \eqref{sol5} increases $\propto r$ and has a profound 
influence on the $\Phi-\chi$ relation. While at small $r$ the dependence 
$\Phi(\chi)$ is a monotonic one within the period $2\pi$, resulting in the single 
valued inverse function $\chi(\Phi)$, in the rings with comparatively large radii 
a nonmonotonic dependence $\Phi(\chi)$ can appear and lead to a multivalued 
dependence of $\chi$ (and, therefore, of $j(\chi)$) on the full magnetic flux $\Phi$.

\begin{figure}[thb!]
\includegraphics[width=0.6\columnwidth,clip=true]{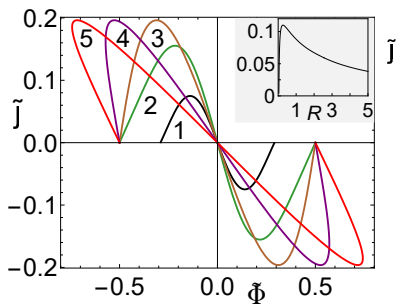}
\caption{$\tilde{\jmath}(\widetilde{\Phi})$
in thin mesoscopic rings closed by $0$ junction with 
$g_\ell=g_\delta=0.1$, at various values of rings's radii:\,\,
(1)\, $R=0.07$\,\,
(2)\, $R=0.16$\,\,
(3)\, $R=1.03$\,\,
(4)\, $R=4.00$\,\,
(5)\, $R=6.50$.
{\it Inset:} The supercurrent $\tilde{\jmath}$, taken at $\widetilde{\Phi}=-0.1$, as a function of the ring's radius.}
\label{fig:jvsPhi}
\end{figure}
The resulting supercurrent-magnetic flux relation for $0$ junction is shown in
Fig.\,\ref{fig:jvsPhi}. The relation between the
magnetic flux and the phase difference for curves 1 and 2
is almost linear, as could be expected from discussing it in 
Sec.~\ref{sec: rmin} in the case of $g_\ell=g_\delta=0.1\ll 1$
and at a sufficiently small $R$. However, the ring's radii that
correspond to curves 4 and 5, are already large enough for the
$v_s$-term to substantially shift the magnetic flux, when the supercurrent
is comparatively large. When $\chi$ is a multiple of $\pi$, the effect
of the $v_s$-term vanishes together with $v_s$ and $j$. This ultimately
results in the multivalued supercurrent-magnetic flux relation, as seen
in curves 4 and 5. Such a behavior implies also a nonmonotonic
radius dependence of the supercurrent at a fixed magnetic flux, as seen in
the inset in Fig.\,\ref{fig:jvsPhi}.

\begin{figure}[thb!]
\includegraphics[width=0.6\columnwidth,clip=true]{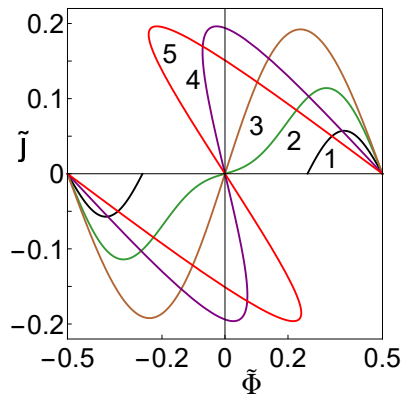}
\caption{$\tilde{\jmath}(\widetilde{\Phi})$
in thin mesoscopic rings closed by $\pi$ junction with
$g_\ell=-0.1$ and $g_\delta=0.3$, at various values of rings's radii:\,\,
(1)\, $R=0.06$\,\,
(2)\, $R=0.10$\,\,
(3)\, $R=0.40$\,\,
(4)\, $R=4.00$\,\,
(5)\, $R=6.50$.}
\label{fig:jvsPhipi}
\end{figure}

The magnetic flux dependence of the supercurrent flowing along the rings
closed by $\pi$ junction, is demonstrated in Fig.\,\ref{fig:jvsPhipi}.
Curve 1 represents the supercurrent-magnetic flux relation for 
the ring's radius satisfying the conditions $R_{\text{min}}(0.5)<R< 
R_{\text{min}}(0)$. Superconductivity is destroyed in such a ring at small 
values of the full magnetic flux $\widetilde{\Phi}$, while it exists, and
a finite supercurrent flows, in a vicinity of $\widetilde{\Phi}=0.5$.

As known, the spontaneous supercurrent and self magnetic flux arise in the 
superconducting rings closed by $\pi$ junction \cite{Bulaevskii1977,%
SigristRice1992,Radovic1999,Hilgenkamp2007}. The spontaneous flux is a 
nontrivial solution of the equation $\Phi=\frac1cLI$, where the total supercurrent 
is considered as a function of the magnetic flux. Since the inductance contribution 
is of importance for the effect, the magnetic-field term $\frac{1}{2c^2}LI^2$ should 
be added to the free energy of the flux-biased ring, identified in Appendix~\ref{sec: sol}. 
A nontrivial solution is, 
as a rule, energetically more favorable than at $\Phi=0$. As the supercurrent vanishes 
at $\Phi=0.5\Phi_0$, for sufficiently large $L$ the solution with a comparatively small 
supercurrent and a flux close to half a flux quantum exists. On the contrary, there is no 
nontrivial solution of the equation $\Phi=\frac1cLI(\Phi)$ for sufficiently small $L$. 
When the difference $R-R_{\text{min}}$ diminishes, the minimum inductance for the nontrivial 
solution to appear increases since the superconductivity region in a vicinity 
of half a flux quantum, as well as the supercurrent within the region, are reduced 
by the inverse proximity effects (see Fig.\,\ref{fig:jvsPhipi}). Therefore, when the 
effects are noticeable and the temperature draws near to $T_c$, the spontaneous 
supercurrent can disappear at sufficiently small nonzero value of $R-R_{\text{min}}$, 
i.e., below $T_c$.

Consider now the evolution of the current-magnetic flux relation with
temperature. The temperature dependence of the supercurrent $j=
\tilde{\jmath}j_{\text{dp}}$ originates
not only from the dimensionless radius $R=\widetilde{R}\tau^{1/2}$ and
from the effective coupling constants $g_{\ell\,(\delta)}=
g_{\ell0\,(\delta0)}/\sqrt{\tau}$, but also from the temperature dependence
of the depairing current $j_{\text{dp}}\propto\tau^{3/2}$. For extracting
the temperature dependence of $j$, it is convenient to switch over to a new
dimensionless quantity $J=j/j_{\text{dp}}^{GL}(0)$, where
$j_{\text{dp}}^{GL}(0)$ is the so called zero-temperature depairing current
of the GL theory $j_{\text{dp}}^{GL}(0)=\tau^{-3/2}j_{\text{dp}}$. The quantity
$j_{\text{dp}}^{GL}(0)$ is known to exceed in $2 - 3$ times the real 
zero temperature depairing current $j_{\text{dp}}(0)$. For example, the equality
$j_{\text{dp}}(0)\approx 0.385j_{\text{dp}}^{GL}(0)$ follows from microscopic
results for the junctions, involving conventional diffusive superconductors 
\cite{Golubov2001}.

\begin{figure}[thb!]
\includegraphics[width=0.75\columnwidth,clip=true]{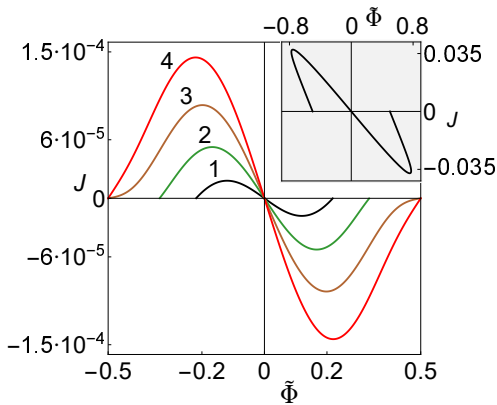}
\caption{$J(\widetilde{\Phi})$ at $\widetilde{R}=5$,\,
$g_{\ell0}=g_{\delta0}=0.1$, taken at various temperatures:\,\,
(1)\, $\tau=0.005$\,\,
(2)\, $\tau=0.006$\,\,
(3)\, $\tau=0.007$\,\,
(4)\, $\tau=0.008$.\,Inset:\,The multivalued
current-magnetic flux relation at $g_{\ell0}=g_{\delta0}=0.1$, $\widetilde{R}=15$ and $\tau=0.1$.
}
\label{fig:jtau32vsPhi}
\end{figure}

The magnetic flux dependence of $J$, taken near $T_c$ at various values of
$\tau$, is shown in Fig.\,\ref{fig:jtau32vsPhi} for
the ring with $\widetilde{R}=5$ and $g_{\ell\,(\delta), 0}=0.1$. The supercurrent takes
quite small values, as it is taken in units of $j_{\text{dp}}^{GL}(0)$, and not of
$j_{\text{dp}}$. At substantially larger values of $\widetilde{R}$ and $J$, the
multivalued current-magnetic flux relation can arise with varying temperature,
and hence $\tau$, as shown in the inset for the case $\tau=0.1\ll1$ and
$\widetilde{R}=15$.

This paper has focused on the mean field results within the GL theory. 
The current-phase relation of the junction in the superconducting ring can also be affected
both by classical fluctuations of the order parameter in or close to the fluctuation region near $T_c$, 
and by quantum fluctuations, which are of importance at very low temperatures. The former case 
is still an open field for further study, while a number of results have already 
been obtained regarding the latter one \cite{Hekking1997}.

\section{Conclusions}
\label{sec: conclusions}

The problem of destructive effects of the Josephson junction on
superconductivity of mesoscopic or nanoscopic rings has been
solved in this paper within the GL theory. The superconducting
state is shown to take place, when the ring's radius exceeds
a minimum radius $r_{\text{min}}$ that depends on the phase difference or the
magnetic flux, as well as on the temperature and the effective
Josephson and interface coupling constants. Depending on the junction
transparency and/or the strength of the pair breaking by the junction interface,
the minimum radius can become a noticeable fraction of the temperature
dependent coherence length, up to $0.5\xi(T)$. The superconductor-normal
metal phase transition that takes place at $r=r_{\text{min}}$, is shown to be
of the second order. The magnetic flux and temperature dependence of $r_{\text{min}}$ allow
an observation of the transition under slowly varying magnetic field or temperature. The
minimum radius increases when the temperature draws near to $T_c$, resulting
in the equality $r=r_{\text{min}}(T_c)$ at the modified critical temperature.

When the ring's radius slightly exceeds $r_{\text{min}}$, not only the
critical temperature of the superconducting state, but also the
current-phase and current-magnetic flux relations can be noticeably modified
by the inverse proximity effects in the ring. The specific features of these 
characteristics have been determined both for $0$ and $\pi$ junctions, and 
their dependence on the ring's radius as well as on the temperature
have been obtained. A substantial evolution of the magnetic flux dependence of 
the Josephson current with increasing radius of thin rings 
has been demonstrated to persist even at large ring's radii and result in a multivalued 
behavior, while the current-phase dependence stays almost unchanged at $r\agt\xi(T)$, 
approaching the one that describes the junctions in asymptotically large rings.
The identified multivaluedness of the supercurrent's magnetic flux dependence
is related to the presence of the corresponding equilibrium and metastable 
states, and with possible transitions between them, which result in the hysteretic 
behavior of the supercurrent. The hysteretic properties, however, lie 
outside the scope of this paper. They appear in the rings with comparatively large radii 
and take place irrespective of the proximity effects' strength, while the paper mainly 
concerns itself with the proximity-induced effects in the rings of smaller sizes.

Currents flowing in individual nanoscopic and mesoscopic superconducting rings can be 
experimentally determined using a number of methods. In addition to methods based on electrical 
transport measurement technique employing direct electrical contacts with the system 
\cite{Liu2001}, there also are the non-invasive ones that use micromechanical torsional 
magnetometers \cite{JGEHarris2013,JGEHarris2016} or measure the ring's susceptibility 
\cite{Moler2007,Moler2011_2}. The present sensitivity and accuracy of the experimental 
techniques as well as modern technological developments for fabrication of superconducting 
nanorings, nanocylinders and nano-SQUIDs \cite{Liu2001,Granata2016} represent 
the basis for possible observations of the theoretical predictions of this paper.

\begin{acknowledgments}
The support of RFBR grant 14-02-00206 is acknowledged.
\end{acknowledgments}

\appendix

\section{Solutions of the GL equations}
\label{sec: sol}

The parameter ${\cal E}$, related to the first integral of 
the GL equation \eqref{gleq2} and defined as
\be
{\cal E}=\left(\dfrac{df(x)}{dx}\right)^2
+\dfrac{i^2}{f^2(x)}+f^2(x)-\dfrac{1}{2}f^4(x),
\label{RPhi88gt}
\ee
is spatially constant, when taken for the solutions 
$f(x)$. 

Taking $x=0$ in \eqref{RPhi88gt} and making use of
\eqref{bcss99} and \eqref{bcss9}, one can express $\cal E$
via $f_0$ and the parameters of the superconducting ring closed by the junction:
\be
{\cal E}=\bigl(1+g_b^2+g_\ell^2\sin^2\chi\bigr)f_0^2-\dfrac{1}{2}f_0^4\,.
\label{calef0}
\ee

Eq. \eqref{RPhi88gt} can be also rewritten in the form
\be
\left(\dfrac{df}{dx}\right)^2=\dfrac{1}{2f^2}(f^2-f_+^2)(f^2-f_d^2)(f^2-f_-^2).
\label{RPhi89gt}
\ee

The quantities $t_-=f_-^2$, $t_d=f_d^2$ and $t_+=f_+^2$ satisfy the
following set of equations
\begin{align}
t_-+t_d+t_+=2,&\qquad t_-t_dt_+=2i^2, \nonumber\\
t_dt_-+t_dt_++&t_-t_+=2{\cal E}.
\label{ttt1}
\end{align}

Apart from the boundary and periodic conditions, the solution of
equation \eqref{RPhi89gt} is characterized by three formal extrema $f_-,
\,f_d,\,f_+$ with vanishing first derivatives $\frac{df}{dx}$.
When the ring closed by the junction is in the equilibrium, only one of
the extrema represents an actual (maximum) value of $f(x)$ in the ring,
while the other two are just auxiliary quantities. Indeed, a
periodic inhomogeneous solution $t(x)=f^2(x)$ with only
a single minimum and a single maximum in the ring, should energetically be
the most favorable one. The minimum $t_0=f_0^2$ is induced
at the junction interface $x=0$ by the pair breaking effects. In
contrast to the extrema described by \eqref{RPhi89gt} and \eqref{ttt1},
the derivative at $x=0$ is nonzero and discontinuous in accordance
with the boundary conditions \eqref{bcss99}. In the equilibrium, one expects
the maximum of $t(x)$ to coincide with one of the roots of the right
hand side of \eqref{RPhi89gt} (let it be $t_d$), and to be realized at the point
$x=\pi R$ diametrically opposed to $x=0$. In general,
either all three roots $t_-,\, t_d$ and $t_+$ take real values, or only
one is real and the other two are the complex conjugate of each other. As the
left hand side of \eqref{RPhi89gt} takes nonnegative values, one concludes
that the minimum that does not actually show up in the ring (let it be
$t_-$), has to be real in both cases. Since one of the other two
extrema is the real quantity $t_d$, both of them also have to be real.

Let $t(x)$ have a minimum $t_0$ at $x=0$ and a maximum
$t_d$ at $x=\pi R$. A nonnegative value of the left hand
side of \eqref{RPhi89gt} entails the existence of one more maximum $t_+$.
Assuming that $t_-\le t_0\le t(x)\le t_d\le t_+$, the
solution of \eqref{RPhi89gt} can be represented in the region $|x|\le\pi R$ as
\begin{multline}
|x|=\sqrt{\dfrac{2}{t_+-t_-}}\left[
F\left(\left.\arcsin\sqrt{\dfrac{t-t_-}{t_d-t_-}}\right|\,\dfrac{t_d-t_-}{t_+-t_-}\right)- \right.\\
\left. -
F\left(\left.\arcsin\sqrt{\dfrac{t_0-t_-}{t_d-t_-}}\right|\,\dfrac{t_d-t_-}{t_+-t_-}\right)\right].
\label{sol1}
\end{multline}
Here $F\left(\varphi\left|\,m\right.\right)$ is the elliptic integral of the first kind.
The notations of arguments of elliptic integrals vary in literature and here the definitions of
the Mathematica book are used \cite{Wolfram2003}. Making use of the addition
theorem for $F(\varphi|m)$ \cite{PrasolovSolovyev1997}, the solution \eqref{sol1} can be rewritten
in the form
\be
|x|=\sqrt{\dfrac{2}{t_+-t_-}}F\left(\psi\left|\,\dfrac{t_d-t_-}{t_+-t_-}\right.\right),
\label{sol2}
\ee
where
\begin{widetext}
\be
\sin\psi=\dfrac{\sqrt{\left(t-t_-\right)\left(t_d-t_0\right)\left(t_+-t_0\right)}-
\sqrt{\left(t_0-t_-\right)\left(t_d-t\right)\left(t_+-t\right)}}{\left(t_d-t_-\right)\left(t_+-t_-\right)-
\left(t-t_-\right)\left(t_0-t_-\right)}\sqrt{t_+-t_-}\enspace .
\label{sol3}
\ee
\end{widetext}

The solution \eqref{sol2}, \eqref{sol3}
contains four parameters $t_-$, $t_d$, $t_+$ and $t_0$. They are linked to each other
by the system of three equations \eqref{ttt1}, where expressions \eqref{bcss9} and
\eqref{calef0} should be substituted for ${\cal E}$ and $i$. The fourth equation is obtained
taking $x=\pi R$ and $t(\pi R)=t_d$ in \eqref{sol2}:
\begin{multline}
\!\pi R=\\ \!=\!
\sqrt{\dfrac{2}{t_+-t_-}}
F\!\left(\!\arcsin\sqrt{\dfrac{(t_+-t_-)(t_d-t_0)}{(t_d-t_-)(t_+-t_0)}}\,\left|\dfrac{t_d-t_-}{t_+-t_-}\right.\!\right).
\label{sol4}
\end{multline}

When the relation \eqref{sol4} is taken into account, the solution \eqref{sol2}-\eqref{sol3} is 
simplified and can be reduced to the form
\begin{multline} 
|x|=\pi R-\\ -\,\sqrt{\dfrac{2}{t_+-t_-}}
F\biggl(\arcsin\sqrt{\dfrac{\left(t_+-t_-\right)\left(t_d-t\right)}{\left(t_d-t_-\right)\left(t_+-t\right)}}
\,\left|\dfrac{t_d-t_-}{t_+-t_-}\right.\biggr), \\ |x|\le\pi R.
\label{tonx4b}
\end{multline}

A joint solution of \eqref{ttt1} and \eqref{sol4} represents $t_-,\,t_0,\,t_d$ and $t_+$,
as well as the whole of the inhomogeneous profile of the order parameter \eqref{tonx4b},
as functions of the phase difference $\chi$ and of the dimensionless radius of the ring
$R=r\big/\xi(T)$.

In the case of a flux-biased ring the relationship \eqref{sol5} between the phase 
difference and the magnetic flux should be used. 
The integration on the right hand side of \eqref{sol5} can be carried
out after inserting the order parameter profile \eqref{tonx4b}.
The result contains the elliptic integral of the third kind and takes the form
\begin{widetext}
\be
\widetilde{\Phi}-n=-\left\{\dfrac{\chi}{2\pi}+\,i\Biggl[\dfrac1{t_+}R
+\dfrac{\sqrt{2}\left(t_+-t_d\right)}{\pi t_+t_d\sqrt{t_+-t_-}}\,\Pi\left(\dfrac{t_+\left(t_d-t_-\right)}{t_d\left(t_+-t_-\right)};
\left.\arcsin\sqrt{\dfrac{\left(t_+-t_-\right)\left(t_d-t_0\right)}{\left(t_d-t_-\right)\left(t_+-t_0\right)}}
\,\right|\dfrac{t_d-t_-}{t_+-t_-}\right)\Biggr]\right\}\, .
\label{sol6}
\ee
\end{widetext}

When the solutions of \eqref{ttt1}, \eqref{sol4} for $t_-$, $t_d$, $t_+$ and $t_0$, along
with the second expression in \eqref{bcss9} for the supercurrent, are inserted on the right
hand side of \eqref{sol6}, the latter represents the quantity $\widetilde{\Phi}-n$
as a function of $\chi$ and $R$. As follows from \eqref{sol6}, the magnetic flux equals an 
integral (a half-integral) number, when the phase difference is an even (odd) multiple of 
$\pi$, since the supercurrent vanishes at $\chi=m\pi$. Since the change of the winding number by one 
is, by definition, the change of the order-parameter phase $\phi(x)$ by $2\pi$ after it has gone 
around the loop, the changes of $\chi$ by a multiple of $2\pi$ and of the winding number by an 
integer are unambiguously interrelated in \eqref{sol6}. The $2\pi$-periodic dependence of $t_-$, 
$t_d$, $t_+$ and $t_0$ on $\chi$ that follows from \eqref{ttt1}, \eqref{sol4} and \eqref{bcss9}, 
\eqref{calef0}, should be noted here. 

If the critical current is small as compared to the depairing current and the ring's radius 
satisfies the condition $R\alt 1$, the quantity $\widetilde{\Phi}-n$ in \eqref{sol6}, as a 
function of $\chi$ and $R$, can be inversed resulting in a single valued function $\chi(\widetilde{\Phi}-n,R)$. 
This allows one to obtain all the quantities as functions of $\widetilde{\Phi}-n$ and $R$. Similar to the 
unbroken rings, the dependence of the winding number on the magnetic flux should be determined from the 
minimization of thermodynamic potential at a fixed $\widetilde{\Phi}$. In this case one gets the physical 
quantities in the equilibrium state as periodic functions of $\widetilde{\Phi}$ with unit period: 
$\chi(\widetilde{\Phi}+1)=\chi(\widetilde{\Phi})+2\pi$.

While the superconducting state in the ring closed
by the junction is inevitably inhomogeneous unless the effective
coupling constants $g_\ell$ and $g_\delta$ vanish, the equilibrium state of the
unbroken cylindrically symmetric thin ring is characterized by the spatially
constant absolute value of the order parameter. However, an inhomogeneous
profile of the order parameter can arise as an unstable (metastable) state of
the uninterrupted rings \cite{Horane1996,Vodolazov2002,CastroLopez2005}.
Such a nonuniform solution follows from \eqref{sol4} - \eqref{sol6} in
the limit $t_0\to t_-$ and $\chi\to0$. \footnote{In more involved cases one should
also take in this limit $g_\delta\to0$,\, $g_\ell\sin\chi\to -\tilde{v}_s(0)$ and
$g_\ell\sin^2\frac{\chi}{2}\to0$, where $v_s=(2|a|^{1/2}K^{1/2}\big/\hbar)
\tilde{v}_s(x)$.} For the unbroken rings, the three extrema $t_-$, $t_d$
and $t_+$ can be calculated, as functions of $\widetilde{\Phi}$ and $R$, based
on the first equation in \eqref{ttt1}, as well as on \eqref{sol4} and \eqref{sol6}.
The inhomogeneous solution does not always exist in the unbroken axially symmetric
rings, for example, at sufficiently small magnetic fluxes, since the condition
$\chi\equiv0$ makes the equation \eqref{sol6} to be substantially more restrictive
than in the case of rings with a junction. The junction breaks the 
ring's  axial symmetry, that modifies the nonuniform solution and stabilizes it. 
Some other examples of its stabilization in rings with the broken symmetry 
have been discussed earlier \cite{Berger1995,Berger1997,Vodolazov2007}.

The spatial integration can also be taken analytically in the expression
for the bulk thermodynamic potential, with the solution
\eqref{tonx4b} of the GL equation inserted in \eqref{F1gt}. As a result,
one gets the free energy in the form
\begin{widetext}
\begin{multline}
\widetilde{\cal F}=g_bt_0-\,
\dfrac{\sqrt{2}}3\sqrt{\bigl(t_{+}-t_0\bigr)\bigl(t_d-t_0\bigr)\bigl(t_0-t_{-}\bigr)}\,
-\dfrac{\pi R}{3}\Bigl[2t_+-\bigl(1+g_b^2+g_\ell^2\sin^2\chi\bigr)t_0+\dfrac{1}{2}t_0^2\Bigr]+\\
+\dfrac{2\sqrt{2}}{3}\biggl[\sqrt{t_+-t_-}\,E\Biggl(\left.\arcsin\sqrt{\dfrac{\left(t_+-t_-\right)
\left(t_d-t_0\right)}{\left(t_d-t_-\right)\left(t_+-t_0\right)}}\,\right|\,\dfrac{t_d-t_-}{t_+-t_-}\Biggr)
-\sqrt{\dfrac{\left(t_d-t_0\right)\left(t_0-t_-\right)}{\left(t_+-t_0\right)}}\biggr]\,,
\label{feL8p}
\end{multline}
\end{widetext}
where $\widetilde{\cal F}=(b{\cal F})\Big/(2K^{1/2}|a|^{3/2})$, ${\cal F}={\cal F}_{b}+
{\cal F}_{\text{int}}$, and $E(\left.\varphi\,\right|m)$ is the elliptic integral of the
second kind. Similar to Eq.~\eqref{sol6} for the magnetic flux, the free energy
\eqref{feL8p} is given as a function of $\chi$ and $R$, as the
quantities $t_-$, $t_d$, $t_+$ and $t_0$ are the solutions of \eqref{ttt1} and
\eqref{sol4}. 

Thermodynamic potential \eqref{feL8p} takes into account both the bulk and the proximity-modified
interface contributions. Near the transition point $R=R_{\text{min}}$ the two contributions strongly 
compete with each other and the result is described by formula \eqref{ttt4} of the paper.

Since a joint solution of equations \eqref{ttt1}, \eqref{sol4} and \eqref{sol6}
allows one to get the supercurrent \eqref{bcss9} as a function of $R$ and of the full magnetic 
flux $\Phi$, the relation
\be
\Phi_e=\Phi-\dfrac1cLI
\label{sol6a}
\ee
gives in this case the applied magnetic flux $\Phi_e(\Phi,R)$. Here $I={\cal A}j$ 
is the total supercurrent, ${\cal A}$ is the cross section's area and $L$ is the 
inductance. As discussed in Sec.~\ref{sec: intro}, the inductance effects can be 
safely ignored whenever $R-R_{\text{min}}$ is sufficiently small. The main focus 
of the paper is on a relatively small $R-R_{\text{min}}$, when the self-field 
effects can be mostly disregarded.

\section{Derivation of minimum radius}
\label{sec: mrc}

An effect of the ring's size $R$ on the inhomogeneous solution of the GL
equation is described by equation \eqref{sol4}.
Were the first argument of the elliptic integral on the right hand side of
\eqref{sol4} arbitrarily small at a finite $t(x)$,
the equation \eqref{sol4} would allow superconductivity in the rings with
very small radii, on the scale of the GL theory. On account of the
conditions $0\le t_-\le t_0\le t(x)\le t_d\le t_+$, the first
argument vanishes only at $t_d=t_0$, i.e., for the uniform 
order parameter. However, such a profile is incompatible with the
proximity-induced boundary conditions \eqref{bcss99} unless the effective
coupling constants $g_\ell$, $g_\delta$ vanish. Indeed, substituting the
equality $t_d=t_0$, as well as \eqref{bcss9} and \eqref{calef0}, in
\eqref{ttt1}, one gets the homogeneous normal metal state, where $t_d=
t_0=t_-=0$ and $t_+=2$.

In order to obtain the minimum radius $R_{\text{min}}$, one should find
not only the limits of the individual extrema $t_d,\,t_-$ and $t_+$ at
the transition ($t_0\to0$), but also of the
combinations of these quantities, which form the arguments of the
elliptic integral in \eqref{sol4}. Small deviations from the individual
limits have to be considered for this purpose. As seen
in \eqref{calef0} and \eqref{bcss9}, both ${\cal E}$ and $i$ are the
small parameters, when the minimum of $t(x)$ is small,
$t_0\ll1$. It is the case, when the radius of a superconducting ring
only slightly exceeds the minimum radius. Up to the first order terms
in ${\cal E}$ and $i$, the solutions of equations \eqref{ttt1} are
\be
t_{d\,(-)}=\dfrac{\cal E}{2}\pm\sqrt{\dfrac{{\cal E}^2}{4}-i^2}, \quad
t_+=2-{\cal E}.
\label{ttt2}
\ee

It follows from \eqref{ttt2}, as well as from the conditions $t_0\le
t(x)\le t_d$, that superconductivity is destroyed throughout
the ring simultaneously, when the minimal value of the order parameter
$f_0$ vanishes. The absence of an isolated phase slip center $f_0=0$
at $x=0$ is a direct consequence of the boundary conditions \eqref{bcss99}. 

Substituting expressions \eqref{calef0} and \eqref{bcss9} for $\cal E$
and $i$ in \eqref{ttt2}, one finds in the limit $t_0\to0$, that the
second argument of the elliptic integral in \eqref{sol4} vanishes
while the first argument remains finite. The elliptic integral of the
first kind coincides with its first argument under such conditions, so
that \eqref{sol4} transforms into Eq.~\eqref{ttt3} for the minimum
radius. 

If the phase difference $\chi$ across the junction is controlled by
the magnetic flux penetrating the superconducting ring, the minimum
radius actually depends on $\widetilde{\Phi}$, rather than on $\chi$.
The relation \eqref{sol6} between the quantities $\chi$ and
$\widetilde{\Phi}$ substantially simplifies at $R=R_{\text{min}}(\chi)$.
Taking the limit $t_0\to0$ in \eqref{sol6}, one can make the substitution
$t_0=t_-=t_d=0$ and $t_+=2$ everywhere except for the extrema combinations,
the calculation of which requires taking account of small deviations from
the presented individual limits. On top of that, the expression \eqref{bcss9}
for the supercurrent and the relation $\Pi(n;\varphi|0)=\frac{1}{\sqrt{1-n}}
\arctan\left(\sqrt{1-n}\tan\varphi\right)$ should be used. The remaining
combinations of quantities $t_d$, $t_0$, and $t_-$ can be found in the limit
$t_0\to0$ using \eqref{ttt2}, \eqref{calef0} and \eqref{bcss9}. As a result,
one obtains Eq.~\eqref{sol8} of the paper.

\providecommand{\noopsort}[1]{}\providecommand{\singleletter}[1]{#1}%

\end{document}